\documentclass{article}

\usepackage[preprint]{neurips_2024}

\usepackage[utf8]{inputenc} 
\usepackage[T1]{fontenc}    
\usepackage{hyperref}       
\usepackage{url}            
\usepackage{booktabs}       
\usepackage{amsfonts}       
\usepackage{nicefrac}       
\usepackage{microtype}      
\usepackage{xcolor}         
\usepackage{multirow}
\usepackage{subfigure}
\usepackage[normalem]{ulem}
\useunder{\uline}{\ul}{}
\usepackage[ruled,linesnumbered, boxed]{algorithm2e}
\usepackage{xspace}
\usepackage{graphicx}
\usepackage{amsmath}
\usepackage{float}
\usepackage{makecell}
\usepackage{color}
\usepackage{tabulary}
\usepackage{mathtools}

\usepackage{amssymb}
\usepackage{url}
\usepackage{diagbox}
\usepackage{bm}
\usepackage{enumitem}
\usepackage{marvosym}
\usepackage{ifsym}

\newcommand{\eg}{\emph{e.g.,}\xspace}
\newcommand{\ie}{\emph{i.e.,}\xspace}

\newcommand{\name}{LATEX-GCL\xspace}

\usepackage{color}

\usepackage{color, xspace}

\title{LATEX-GCL: Large Language Models (LLMs)-Based Data Augmentation for Text-Attributed Graph Contrastive Learning}

\author{%
  Haoran Yang\\
  University of Technology Sydney \& \\
  The Hong Kong Polytechnic University \\
  \texttt{haoran.yang-2@student.uts.edu.au} \\
  \And
  Xiangyu Zhao\thanks{Corresponding author.} \\
  City University of Hong Kong \\
  \texttt{xianzhao@cityu.edu.hk} \\
  \And
  Sirui Huang \\
  The Hong Kong Polytechnic University \& \\
  University of Technology Sydney \\
  \texttt{sirui.huang@connect.polyu.hk} \\
  \And
  Qing Li\footnotemark[1] \\
  The Hong Kong Polytechnic University \\
  \texttt{qing-prof.li@polyu.edu.hk} \\
  \And
  Guandong Xu\footnotemark[1] \\
  The Education University of Hong Kong \& \\
  University of Technology Sydney \\
  \texttt{gdxu@eduhk.hk} \\
}

\begin{document}

\maketitle

\begin{abstract}
Graph Contrastive Learning (GCL) is a potent paradigm for self-supervised graph learning that has attracted attention across various application scenarios. However, GCL for learning on Text-Attributed Graphs (TAGs) has yet to be explored. Because conventional augmentation techniques like feature embedding masking cannot directly process textual attributes on TAGs. A naive strategy for applying GCL to TAGs is to encode the textual attributes into feature embeddings via a language model and then feed the embeddings into the following GCL module for processing. Such a strategy faces three key challenges: I) failure to avoid information loss, II) semantic loss during the text encoding phase, and III) implicit augmentation constraints that lead to uncontrollable and incomprehensible results. In this paper, we propose a novel GCL framework named \name to utilize Large Language Models (LLMs) to produce textual augmentations and LLMs' powerful natural language processing (NLP) abilities to address the three limitations aforementioned to pave the way for applying GCL to TAG tasks. Extensive experiments on four high-quality TAG datasets illustrate the superiority of the proposed \name method. The source codes and datasets are released to ease the reproducibility, which can be accessed via this link\footnote{https://anonymous.4open.science/r/LATEX-GCL-0712}.
\end{abstract}

\section{Introduction}
\label{sec:intro}
In numerous real-world scenarios, graph data is often enriched with textual attributes, for instance, user-item interaction graphs in recommendation systems that include textual user profiles and product descriptions \cite{amazon-1, amazon-2}. This type of graph data is referred to as TAGs \cite{llm-survey}. More than recommendation systems, the application scenarios of TAGs also include bioinformatics \cite{bio-info}, computer vision \cite{cv}, and quantum computing \cite{quantum}. The development of effective methodologies for processing and analyzing TAGs is crucial for advancing applications that rely on such data. With the advent of graph learning techniques, a variety of paradigms have been introduced. Notably, GCL \cite{dgi, mvgrl, dsgc} has gained prominence as a powerful self-supervised technique for graph representation learning, capitalizing on the benefits of self-supervision in cases of lacking sufficient labels. Current GCL approaches typically employ perturbations to manipulate graph structures and feature embeddings, thereby generating contrasting samples for GCL \cite{graphcl, gca, grace, cgc}. Despite the diversity of these strategies, they fall short in directly augmenting the textual attributes inherent in TAGs. Consequently, there is a pressing need to devise a framework that synergizes GCL with TAGs, potentially enhancing the performance of graph learning tasks within TAG application scenarios by harnessing the strengths of GCL techniques.

Despite the advancements in GCL, the literature reveals a gap in the development of GCL methodologies specifically tailored for TAG settings \cite{llm-survey, llm-graph-survey, dataset}. An initial attempt to address this, referred to as Topological Contrastive Learning (TCL) for TAGs, is outlined in \cite{dataset}. This approach begins by encoding textual attributes into feature embeddings for each node. Subsequently, it employs conventional GCL augmentations such as feature masking and proximity perturbation \cite{graphcl} to process the graph, followed by the execution of the remaining GCL steps in sequence. While this rudimentary approach enables the adaptation of GCL to TAG settings, it is not without significant drawbacks that could potentially compromise its effectiveness. There are three limitations lie ahead: 
I) \textbf{Information Loss.} Existing research \cite{mvgrl} has identified information loss as a significant issue during the augmentation phase of conventional GCL methods, attributable to randomness and noise inherent in these processes. Adhering to the aforementioned rudimentary pipeline and employing standard random-based augmentation techniques, such as feature masking, inevitably leads to this loss of information. To enhance the performance of graph models within the GCL framework, it is imperative to implement strategies that mitigate such information loss. 
II) \textbf{Incapable Language Models.} The encoding of textual attributes in TAGs presents challenges when using both shallow text embedding methods, such as bag-of-words \cite{bow} and skip-gram \cite{skip-gram}, and advanced deep language models like BERT \cite{bert}, DeBERTa \cite{deberta}, and GPT-2 \cite{gpt-2}. Shallow embedding methods are constrained by their limited capacity to capture nuanced semantic features, whereas deep language models, despite their sophistication, fall short in complex reasoning tasks \cite{llm-survey}. The reliance on these inadequate language models for the text encoding phase leads to an inevitable semantic degradation contained in the original textual attributes. 
III) \textbf{Implicit Constraint on Augmentations.} Conventional GCL methods \cite{dgi, graphcl}, as well as those employing sophisticated adaptive augmentation strategies \cite{grace, cgc}, share a fundamental challenge: the absence of explicit constraints on the augmentation process. This deficiency hinders users from monitoring and comprehending the effects of augmentation techniques, leading to augmented outcomes that are both uncontrollable and incomprehensible.

To overcome the aforementioned limitations, we introduce a novel approach named LATEX-GCL that employs an LLM to generate auxiliary texts, which act as augmented textual attributes for GCL applied to TAGs. This method circumvents the information loss associated with conventional feature augmentation techniques (\eg random feature masking). Thanks to the general knowledge contained in the LLM \cite{llm-knowledge-base}, our strategy effectively enriches the semantics of the original text via the LLM-based augmentation, compensating for potential semantic deficits incurred during the text encoding phase. Furthermore, the utilization of LLMs involves natural language inputs, carefully crafted prompts to steer the augmentation process, and outputs that are inherently understandable for human beings. This process ensures that the augmentation constraints and results are explicit and comprehensible, enhancing the transparency and control over the augmentation procedure.
Nevertheless, employing LLMs for textual attribute augmentation in GCL is challenging, as there is a dearth of precedents in the literature to guide such an application. In this paper, we seminally propose a suite of prompts for textual attribute augmentation using LLMs, drawing inspiration from the foundational principles of conventional graph augmentations as cataloged in GraphCL \cite{graphcl}, including \textit{shorten}, \textit{rewriting}, and \textit{expansion}, to facilitate the LLM-based textual attribute augmentation process.

In summary, to address the limitations in current methods and better adapt GCL techniques to TAG settings, we: I) propose a novel GCL framework that can leverage the advantages of LLMs to conduct textual attribute augmentation, II) seminally summarize three types of LLM-based textual attribute augmentations and list the related prompt designs, and III) conduct comprehensive experiments to illustrate the performance and verify the effectiveness of the proposed LATEX-GCL method.

\section{Methodology}
\begin{figure*}[ht]
	\centering
	\includegraphics[width=0.94\textwidth]{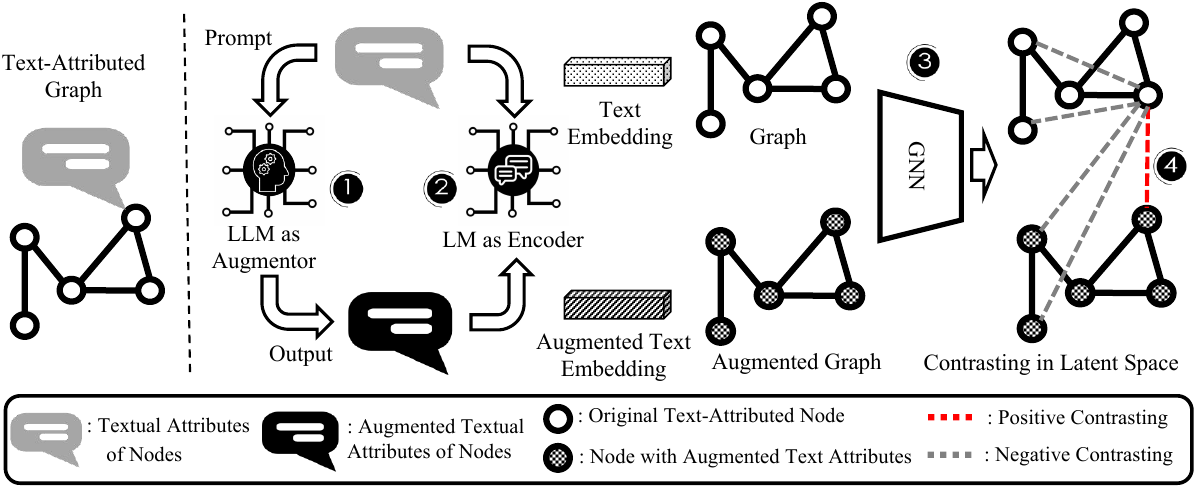}
        \caption{The overview of the proposed method \name.}
	\label{fig:overview}
\end{figure*}

This section illustrates the details of the \name method, starting with the preliminaries, followed by the descriptions for each module, including I) LLM-based text feature augmentation, II) text attribute encoding, III) graph encoding, and IV) graph contrastive learning, as shown in Figure \ref{fig:overview}.

\subsection{Preliminaries and Notations}
Before giving detailed descriptions of the proposed method, some necessary notations and formulations related to TAGs, LLMs, the text encoder, and the graph encoder are listed in this part.

\textbf{Text-Attributed Graphs.} Technically, a TAG can be defined as $\mathcal{G}=(\mathcal{V}, \mathcal{E}, \{t_n\}_{n\in\mathcal{V}})$, where $\mathcal{V}$ is the set of all nodes, $\mathcal{E}$ is the set of all existing links between the nodes in $\mathcal{V}$, and $t_n$ is a sequence of text attributes associated with the $n$-th node. To facilitate the presentation of a graph, an adjacency matrix $\mathbf{A}\in\{0, 1\}^{N\times N}$, where $N$ is the number of nodes, is adopted to demonstrate nodes and links.

\textbf{Large Language Models as Augmentor.} In the proposed method, an LLM is applied as an augmentor to augment the original text attributes in the given TAG guided by the properly designed prompt. In this paper, we use the $LLM(\cdot)$ to denote this augmentor. 
Given the original text attribute $t_n$ and the prompt $p$, we can have the prompted text attribute $\hat{t}_n$. The augmentor $LLM(\cdot)$ finally takes the prompted text attribute $\hat{t}_n$ to output $o_n$.

\textbf{Text Attribute Encoder.} To facilitate the utilization of the original and the augmented text attributes, a text encoder, such as BERT \cite{bert} and DeBERTa \cite{deberta}, is required to obtain feature embeddings. In this paper, $LM(\cdot)$ is used to denote the text encoder, which takes the original text attribute $t_n$ or the augmented text attribute $o_n$ as the input to produce feature embedding $h_n$. Then, the feature embeddings of all the nodes are concatenated to construct the feature matrix $\mathbf{H}$. 

\textbf{Graph Encoder.} A GNN model, such as GCN \cite{gcn}, is implemented to serve as the graph encoder to capture the graph structure information. The graph encoder takes the adjacency matrix and the feature matrix as the inputs to update the feature matrix iteratively, where $g(\cdot, \cdot)$ denotes the graph encoder. A $K$-layer graph encoder will output $\mathbf{H}^{(K)}$ as the final feature embedding matrix.

\subsection{Large Language Model-Based Text Feature Augmentation}
\label{sec:aug}

An LLM is adopted in our proposed method \name as an augmentor to conduct augmentations on the original textual attributes in the input TAG. Adopting the LLM aims to effectively address the three limitations in the aforementioned rudimentary TCL strategy \cite{dataset} in the introduction section, including information loss, incapable language models, and implicit constraints on the augmentation process. However, the adoption of the LLM is non-trivial. A dearth of precedents in the current literature guides how to prompt the LLM to acquire proper augmented texts for the following GCL procedures. In this section, we innovatively propose and summarize a suite of prompts in order to employ the LLM to conduct textual attribute augmentations tailored for GCL on TAGs. 

\begin{table}[h]
\caption{Augmentation strategies for text attribute augmentation.}
\label{tab:aug}
\resizebox{\textwidth}{!}{
\begin{tabular}{c|l|l}
\hline
\textbf{Augmentation} & \multicolumn{1}{c|}{\textbf{Prompt Design}} & \multicolumn{1}{c}{\textbf{Underlying Prior}} \\ \toprule
Shorten                 & \makecell[l]{\textbf{Request:} The following content is the description of \{XXX\}. \textit{Please simplify}\\\textit{and summarize the provided content in one short sentence.}\\\textbf{Content:} \{......\}}                              & \makecell[l]{The shorten augmentation can help\\filter out the redundant contents and\\maintain the key information.}                                           \\ \midrule
Rewriting                  & \makecell[l]{\textbf{Request:} The following content is the description of \{XXX\}. \textit{Please rewrite}\\\textit{the provided content to improve the spelling, grammar, clarity, concision,}\\\textit{logical coherence, and overall readability.}\\\textbf{Content:} \{......\}}                                & \makecell[l]{The rewriting augmentation can help\\identify the invariant semantics contained\\in the original texts.}                                           \\ \midrule
Expansion                  & \makecell[l]{\textbf{Request:} The following content is the description of \{XXX\}. \textit{Please expand}\\\textit{the provided content to give more related and necessary information.}\\\textbf{Content:} \{......\}}                                & \makecell[l]{The expansion augmentation can help\\introduce auxiliary information to enrich\\the original text features.}                                           \\ \bottomrule
\end{tabular}
}
\end{table}

The paradigm of the LLM is known as `pre-train, prompt, and output' \cite{llm-survey}, which is different from the existing language models. An LLM is normally trained on large-scale text corpora and possesses massive general knowledge \cite{llm-survey, llm-knowledge-base}. A properly designed prompt is required to help the LLM output the desired content from the massive knowledge. The prompt has various forms, such as several words or a sentence, and can include additional information to guide and constrain the output of the LLM \cite{tape}. Formally, let $t_n$ be the original text attributes of a node and $p$ denote the prompt to be placed in front of $t_n$, the prompted textual attributes after tokenization can be formalized as $\hat{t}_n=(p_1, p_2, \cdots, p_a, t_{n,1}, t_{n,2}, \cdots, t_{n, b})$
. The LLM-based augmentor $LLM(\cdot)$ is trained to assign a probability to each possible output $o_n=(o_{n,1}, o_{n,2}, \cdots, o_{n, c})$ that consists of $c$ tokens, where the most satisfactory output is expected to have the largest probability value. The probability of the output $o$ given $t_n$ can be formalized as:
\begin{equation}
    p(o_n|\hat{t}_n)=\prod\limits_{i=1}^bp(o_{n,i}|o_{n,<i}, \hat{t}_n).
\end{equation}

To guide the LLM-based augmentor $LLM(\cdot)$ to adapt to the scenario of text-attributed graph contrastive learning, three general text augmentations are proposed, which are listed in Table \ref{tab:aug}. The related discussions about the intuitive priors behind these augmentations are shown below:

\textbf{Shorten.} Given an original text attribute $t_n$, the \textit{shorten} augmentation applies a prompt $p^s$ to produce $\hat{t}^s_n$ to guide $LLM(\cdot)$ output $o^s_n$. Such an augmentation aims to simplify the original text attribute. The underlying prior enforced by it is that simplified content can help filter out redundant information and maintain the key points in the original text attribute.

\textbf{Rewriting.} Given an original text attribute $t_n$, the \textit{rewriting} augmentation applies a prompt $p^r$ to produce $\hat{t}^r_n$ to guide $LLM(\cdot)$ output $o^r_n$. Such an augmentation aims to rewrite the original text attribute so that the invariant semantics contained in the original text attribute can be identified. Moreover, the readability can also be improved to produce high-quality feature embeddings.

\textbf{Expansion.} Given an original text attribute $t_n$, the \textit{expansion} augmentation applies a prompt $p^e$ to produce $\hat{t}^e_n$ to guide $LLM(\cdot)$ output $o^e_n$. Such an augmentation aims to expand the original text attribute to introduce more related and necessary information to leverage the advantages of the knowledge base
, which is trained on a large volume of the corpus.

Without loss of generality, we take the \textit{shorten} augmentation, denoted by superscript $s$, only to describe the workflow of the proposed method \name in the methodology section and omit the two other augmentations. Formally, we prompt the original text attribute $t_n$ of the $n$-th node in the TAG $\mathcal{G}$ to obtain the prompted input $\hat{t}^s_n$ for the augmentor $LLM(\cdot)$ to have:
\begin{equation}
    o^s_n=LLM(\hat{t}^s_n).
\end{equation}
The operations above repeat on each node in the original TAG $\mathcal{G}$ to have the set of augmented text attributes $\{o^s_n|n\in\mathcal{V}\}$. Finally, we can have the augmented TAG $\mathcal{G}^s=(\mathcal{V}, \mathcal{E}, \{o^s_n\}_{n\in\mathcal{V}})$. 

\subsection{Text Attribute Encoding}
\label{sec:text-encoder}
The proposed method \name applies an LLM to directly augment the original text attributes to produce augmented text attributes instead of adopting the feature masking augmentation, which is one of the conventional graph augmentations \cite{graphcl}. Though, as introduced in the introduction section, the adopted strategy can reduce information loss and leverage the advantages of the LLM's superior semantic comprehension capability, the augmented text attributes are in the form of natural language that cannot be processed by the following graph encoding module (\ie the GNN model)
. Therefore, we adopt a text encoder in the proposed method to encode the original and the augmented text attributes to acquire feature embeddings to facilitate the following procedures.

A relatively small language model, such as BERT \cite{bert} and DeBERTa \cite{deberta}, is adopted to serve as the text encoder because they are more powerful than those conventional text embedding methods \cite{bow, skip-gram} and more efficient than the LLMs. Following the LLM-based augmentation phase, the text encoder $LM(\cdot)$ takes the original and the augmented text attributes to produce the original and augmented feature embeddings, which are shown as follows:
\begin{equation}
    h_n=LM(t_n)\in\mathbb{R}^{d\times1},~h^s_n=LM(o^s_n)\in\mathbb{R}^{d\times1},
\end{equation}
where $d$ is the size of feature embeddings. Then, the feature matrix of the original TAG $\mathcal{G}$ and the augmented TAG $\mathcal{G}^s$ can be acquired as follows:
\begin{equation}
    \mathbf{H}=[h_1; h_2; \cdots; h_N]^T\in\mathbb{R}^{N\times d},~\mathbf{H}^s=[h^s_1; h^s_2; \cdots; h^s_N]\in\mathbb{R}^{N\times d}.
\end{equation}
The feature matrices obtained above can cooperate with the adjacency matrix $\mathbf{A}$ of the input TAG to facilitate the following graph encoding procedures.

To enhance performance, it is common to train the text encoder in conjunction with subsequent modules, yet this approach demands substantial computational resources. In practical applications, an adaptor module, typically a straightforward neural network component such as a linear layer, is employed to refine the text encoder's output, thereby boosting performance without incurring the costs associated with fine-tuning. Nevertheless, optimizing the adaptor module often necessitates ample supervised training data from specific downstream tasks. The efficacy of the adaptor module within the context of GCL in this paper remains an open question. We investigate this issue in Section \ref{sec:adaptor}, where we examine the impact of the adaptor module on the performance of \name.

\subsection{Graph Encoding}
TAGs contain a rich repository of information. In addition to the previously mentioned textual attribute information, graph structure is also essential for the graph learning tasks on TAGs. Encoding only the text features is insufficient for acquiring comprehensive graph representation, necessitating the adoption of GNN models (e.g., GCN \cite{gcn}) to learn the structural information in the graph. 

Given the feature matrices $\mathbf{H}$ and $\mathbf{H}^s$ 
obtained in the previous text encoding module, the adjacency matrix $\mathbf{A}$, and a $K$-layer graph encoder $g(\cdot, \cdot)$, we can have the updated feature matrices that possess graph structure information as follows:
\begin{equation}
    \mathbf{H}^{(K)}=g(\mathbf{A}, \mathbf{H})\in\mathbb{R}^{N\times d},~{\mathbf{H}^s}^{(K)}=g(\mathbf{A}, \mathbf{H}^s)\in\mathbb{R}^{N\times d}.
\end{equation}
Each layer of the graph encoder functions as a message passing and aggregation process, collecting information from neighboring nodes and updating the node feature embeddings accordingly. 

\subsection{Graph Contrastive Learning}
Typically, TAGs possess extensive text attributes to describe the nodes. However, in real-world scenarios, label sparsity is a common and unavoidable issue, making it infeasible to manually label each node in the TAG due to the prohibitive costs involved. To broaden the applications of TAGs, it is vital to investigate how to employ self-supervised learning paradigms to obtain high-quality graph embeddings from TAGs without label information. GCL has demonstrated the powerful capability to conduct self-supervised graph learning, to this end, being a viable option for the self-supervised learning paradigm on TAGs. 
This section utilizes a GCL module to process the LLM-augmented graphs, finalizing the workflow of the proposed \name method.

A rough GCL setting is revealed in the fourth part of Figure \ref{fig:overview}. During the training, the node embeddings are usually processed in a mini-batch manner. We use $\mathcal{V}_b$ to denote the set of nodes in a training batch. Formally, suppose that the $i$-th node $i\in\mathcal{V}_b$ is the target. 
The original feature embedding of the target and the augmented feature embedding can be obtained as follows:
\begin{equation}
    h_i^{(K)}={\mathbf{H}_{i,:}^{(K)}}^T\in\mathbb{R}^{d\times1},~{h_i^s}^{(K)}={{\mathbf{H}^s_{i,:}}^{(K)}}^T\in\mathbb{R}^{d\times1}
\end{equation}
The two feature embeddings mentioned above originate from the same target node, thus they are expected to exhibit a high degree of similarity. Therefore, we treat such a pair of embeddings as positive contrasting samples. Then, a subset $\mathcal{V}_M\subseteq\mathcal{V}_b\setminus i$ of nodes, where $|\mathcal{V}_b|=M$, is randomly sampled from the mini-batch to collaborate with the original feature embedding of the target, generating $2M$ negative contrasting samples. The negative contrasting sample's original feature embedding and its LLM-augmented embedding are denoted by $\{h_j^{(K)}|j\in\mathcal{V}_M\}$ and $\{{h_j^s}^{(K)}|j\in\mathcal{V}_M\}$
. A similarity function $sim(\cdot, \cdot)$ is adopted to measure the distance between two feature embeddings. Then, InfoNCE \cite{infonce} is adopted as the loss function for the GCL training:
\begin{equation}
    \mathcal{L}=-\log\frac{e^{sim(h_i^{(K)}, {h_i^s}^{(K)})/\tau}}{e^{sim(h_i^{(K)}, {h_i^s}^{(K)})/\tau}+\sum^M_{j\in\mathcal{V}_M}(e^{sim(h_i^{(K)}, h_j^{(K)})}+e^{sim(h_i^{(K)}, {h_j^s}^{(K)})})},
\end{equation}
where $\tau$ denotes the temperature hyperparameter. After the GCL training, the feature matrix $\mathbf{H}^{(K)}$ is updated, and we can obtain the final feature matrix $\mathbf{H}_{final}$ for downstream inference and evaluation.


\section{Experiment}
To demonstrate the effectiveness and the performance of the proposed \name method, we conduct extensive experiments and show the results with insightful analysis in this section. The related experimental settings are also provided in this section.

\subsection{Experimental Settings}
Four TAG datasets collected by \cite{dataset} are selected for experiments in this paper, including Books-Children, Books-History, Ele-Computers, and Ele-Photo, which are extracted from the Amazon dataset \cite{amazon-1, amazon-2}. The statistics and the content of the raw text of each dataset are listed in Table \ref{tab:dataset}. 

Besides the datasets, five impactful GCL methods are selected as baselines for the comparison study, including GraphCL \cite{graphcl}, GCA \cite{gca}, GRACE \cite{grace}, BGRL \cite{bgrl}, and GBT \cite{gbt}. 

More detailed descriptions of datasets and baselines can be found in \textbf{Appendix \ref{app:data-baselines}}. For better reproducibility, \name implementation details and the related evaluation protocols are provided, listed in \textbf{Appendix \ref{app:implementation}} and \textbf{Appendix \ref{app:evaluation}} for reference.

\begin{table}[h]
\centering
\caption{Dataset Statistics}
\label{tab:dataset}
\begin{tabular}{c|ccc|c}
\hline\hline
Dataset        & \#Node & \#Edge    & \#Class & Raw Text Content  \\ \hline
Books-Children & 76,875 & 1,554,578 & 24      & Book Introduction \\
Books-History  & 41,551 & 358,574   & 12      & Book Introduction \\
Ele-Computers  & 87,229 & 721,081   & 10      & Consumer Review   \\
Ele-Photo      & 48,362 & 500,928   & 12      & Consumer Review   \\ \hline\hline
\end{tabular}
\end{table}

\subsection{Experiment Results \& Analysis}
This section lists the experiment results, including the comparison study, the ablation study, and the adaptor module experiment, which are accompanied by detailed and insightful analyses.

\begin{table}[]
\centering
\caption{A comparison study between the baselines and different settings of the proposed \name method, where the figures underlined denote the best performance achieved by baselines, the figures in boldface represent the best result among all methods, and `OOM' indicates that the method is out of the memory when performing on the specific dataset. The suffixes of \name, including (S), (R), and (E), denote different augmentation prompts used for the experiment, which are \textit{shorten}, \textit{rewriting}, and \textit{expansion}, respectively.}
\label{tab:comparison}
\renewcommand\arraystretch{1.5}
\resizebox{\textwidth}{!}{
\begin{tabular}{c|cccc|cccc}
\hline\hline
Dataset                        & \multicolumn{4}{c|}{Books-Children}                                                                           & \multicolumn{4}{c}{Books-History}                                                                             \\ \hline
\diagbox{Methods}{Metrics} & Accuracy (\%)                  & Precision (\%)                 & Recall (\%)                    & F1 (\%)                        & Accuracy (\%)                  & Precision (\%)                 & Recall (\%)                    & F1 (\%)                        \\ \hline
GraphCL                        & 33.87 (std 0.87)          & 11.63 (std 0.96)          & 6.92 (std 0.28)           & 5.94 (std 0.34)           & 72.42 (std 0.52)          & 22.83 (std 0.49)          & 20.64 (std 0.70)          & 20.86 (std 0.64)          \\
GCA                            & 37.23 (std 0.91)          & 20.15 (std 1.07)          & 9.93 (std 0.24)           & 10.87 (std 0.29)          & 72.87 (std 0.63)          & 27.58 (std 0.61)          & 22.07 (std 0.75)          & 22.94 (std 0.69)          \\
GRACE                          & OOM                       & OOM                       & OOM                       & OOM                       & {\ul 77.53 (std 0.58)}    & {\ul 34.34 (std 2.32)}    & {\ul 24.85 (std 0.83)}    & {\ul 26.01 (std 0.72)}    \\
BGRL                           & {\ul 37.99 (std 0.81)}    & 28.16 (std 2.03)          & 12.73 (std 0.22)          & 13.08 (std 0.30)          & 75.36 (std 0.49)          & 30.02 (std 2.24)          & 23.73 (std 0.92)          & 23.97 (std 0.83)          \\
GBT                            & 36.98 (std 0.83)          & {\ul 28.77 (std 1.59)}    & {\ul 13.09 (std 0.18)}    & {\ul 14.01 (std 0.27)}    & 74.97 (std 0.42)          & 31.17 (std 3.42)          & 23.35 (std 0.87)          & 25.13 (std 0.79)          \\ \hline
LATEX-GCL (S)                  & 38.71 (std 0.65)          & 27.86 (std 2.62)          & 11.89 (std 0.27)          & 12.40 (std 0.43)          & 78.65 (std 0.69)          & 32.58 (std 4.47)          & 25.91 (std 0.77)          & 25.55 (std 0.56)          \\
LATEX-GCL (R)                  & 39.30 (std 0.56)          & 28.07 (std 1.14)          & 12.70 (std 0.10)          & 13.38 (std 0.21)          & 79.08 (std 0.65)          & 35.55 (std 7.17)          & 26.98 (std 0.81)          & 27.02 (std 0.73)          \\
LATEX-GCL (E)                  & \textbf{41.72 (std 0.45)} & \textbf{31.27 (std 2.52)} & \textbf{15.50 (std 0.21)} & \textbf{16.81 (std 0.11)} & \textbf{79.22 (std 0.61)} & \textbf{37.28 (std 5.17)} & \textbf{27.31 (std 0.89)} & \textbf{27.51 (std 0.84)} \\ \hline\hline
Dataset                        & \multicolumn{4}{c|}{Ele-Computers}                                                                            & \multicolumn{4}{c}{Ele-Photo}                                                                                 \\ \hline
\diagbox{Methods}{Metrics} & Accuracy (\%)                  & Precision (\%)                 & Recall (\%)                    & F1 (\%)                        & Accuracy (\%)                  & Precision (\%)                 & Recall (\%)                    & F1 (\%)                        \\ \hline
GraphCL                        & 33.48 (std 0.23)          & 35.77 (std 5.37)          & 15.44 (std 2.65)          & 13.79 (std 0.36)          & 42.24 (std 0.45)          & 36.78 (std 8.00)          & 8.97 (std 0.18)           & 6.21 (std 0.32)           \\
GCA                            & 40.79 (std 0.63)          & 47.23 (std 4.23)          & 21.99 (std 1.96)          & 24.67 (std 0.58)          & 45.74 (std 0.27)          & 40.39 (std 7.59)          & 15.61 (std 0.21)          & 14.95 (std 0.19)          \\
GRACE                          & OOM                       & OOM                       & OOM                       & OOM                       & {\ul 55.65 (std 0.37)}    & {\ul 69.37 (std 1.87)}    & {\ul 29.56 (std 0.73)}    & {\ul 33.97 (std 1.17)}    \\
BGRL                           & 44.36 (std 0.61)          & {\ul 49.78 (std 1.39)}    & 28.43 (std 2.11)          & {\ul 32.27 (std 0.54)}    & 53.77 (std 0.40)          & 68.73 (std 2.39)          & 28.88 (std 0.69)          & 32.74 (std 0.95)          \\
GBT                            & {\ul 45.31 (std 0.59)}    & 49.12 (std 2.03)          & {\ul 29.59 (std 1.05)}    & 31.97 (std 0.48)          & 54.68 (std 0.49)          & 67.56 (std 1.59)          & 29.02 (std 0.84)          & 32.93 (std 1.07)          \\ \hline
LATEX-GCL (S)                  & 48.87 (std 0.56)          & 52.60 (std 1.62)          & 29.48 (std 0.38)          & 31.50 (std 0.41)          & 56.54 (std 0.40)          & \textbf{71.48 (std 1.64)} & 29.14 (std 0.92)          & 35.10 (std 1.31)          \\
LATEX-GCL (R)                  & \textbf{50.80 (std 0.51)} & 52.49 (std 1.16)          & \textbf{31.55 (std 0.41)} & \textbf{33.89 (std 0.50)} & \textbf{57.73 (std 0.16)} & 69.64 (std 0.70)          & \textbf{30.77 (std 0.68)} & \textbf{37.14 (std 0.96)} \\
LATEX-GCL (E)                  & 47.24 (std 0.55)          & \textbf{53.26 (std 1.81)} & 27.58 (std 0.33)          & 28.99 (std 0.30)          & 56.39 (std 0.30)          & 70.88 (std 1.11)          & 28.35 (std 0.52)          & 33.86 (std 0.72)          \\ \hline\hline
\end{tabular}}
\end{table}

\subsubsection{Comparison Study}
The results of the comparison study are listed in Table \ref{tab:comparison}, demonstrating the performance of the proposed \name method and the selected baselines regarding the node classification task on the graph. According to the results, we have the following three findings:

\begin{itemize}[leftmargin=*]
    \item Generally, the proposed \name method achieves the best performance in the comparison study among all datasets compared to the selected baselines. Such an observation verifies the effectiveness and the superiority of our proposed \name method. For different augmentation settings, the results reflect a clear pattern. Specifically, \name equipped with \textit{expansion} augmentation performs better on the two Amazon-Books datasets, and \name equipped with \textit{rewriting} augmentation performs better on the two Amazon-Electronics datasets.
    \item The differences among the performance of different augmentation settings of \name are largely due to the difference in the raw text content of the two types of datasets. As listed in Table \ref{tab:dataset}, the raw text content in the book datasets is the book introduction, and that of the electronic datasets is the consumer review. The book introduction usually contains the correct title of the book, which can help the LLM prompted by the \textit{expansion} augmentation to produce informative content that is highly related to the specific book as the augmented textual attributes, which can significantly benefit the following GCL. However, the consumer reviews of the electronic datasets are normally short and neglect to list the full name of the product reviewed. Such textual attributes prevent the LLM prompted by \textit{expansion} augmentation from producing informative content. Even worse, it may lead the LLM to introduce more noise (\ie unrelated content). Therefore, utilizing the LLM to extract key information in the consumer review would be more suitable instead of producing auxiliary information. The experiment results confirm our analysis. On dataset Books-Children and Books-History, \name equipped with \textit{shorten} augmentation and \textit{rewiriting} augmentation, which are both helpful for key information extraction from the original textual attributes as discussed in Section \ref{sec:aug}, outperform \name equipped with \textit{expansion} augmentation. Moreover, in the scenarios of lacking sufficient computational resources, the \textit{shorten} augmentation would be a promising alternative for the \textit{rewriting} expansion as the gap between the performance of these two augmentations is insignificant on both electronic datasets.
    \item GraphCL has the lowest scores across all metrics and datasets. This is because GraphCL uses classical augmentation techniques to conduct GCL, outperformed by those adaptive augmentation strategies. GCA adopts an automatic selection strategy to pick conventional augmentations used in GraphCL, slightly improving the performance. GRACE proposes an adaptive strategy to augment the graph according to the specific input data. However, such a strategy significantly increases the complexity. Consequently, GRACE is out of memory when performing on the two large datasets, including Books-Children and Ele-Computers. The significant improvement brought by the adaptive augmentation strategy is reflected by GRACE's performance on Books-History and Ele-Photo. Specifically, GRACE achieved the best results among all the baselines on these two datasets. Both BGRL and GBT methods follow the same idea of utilizing different training objectives instead of InfoNCE to eliminate the requirement of negative contrasting samples and achieve better performance. We can observe that both methods can perform well on large datasets. However, on the relatively small datasets where GRACE can function, BGRL and GBT are outperformed by GRACE due to both methods taking the same conventional augmentation techniques as adopted by GraphCL, which is less advanced compared to the adaptive augmentation strategy. 
\end{itemize}

\begin{figure*}[h]
	\centering
	\subfigure[Accuracy scores of different settings]{
		\begin{minipage}[t]{0.45\linewidth}
			\centering
			\includegraphics[width=1\textwidth]{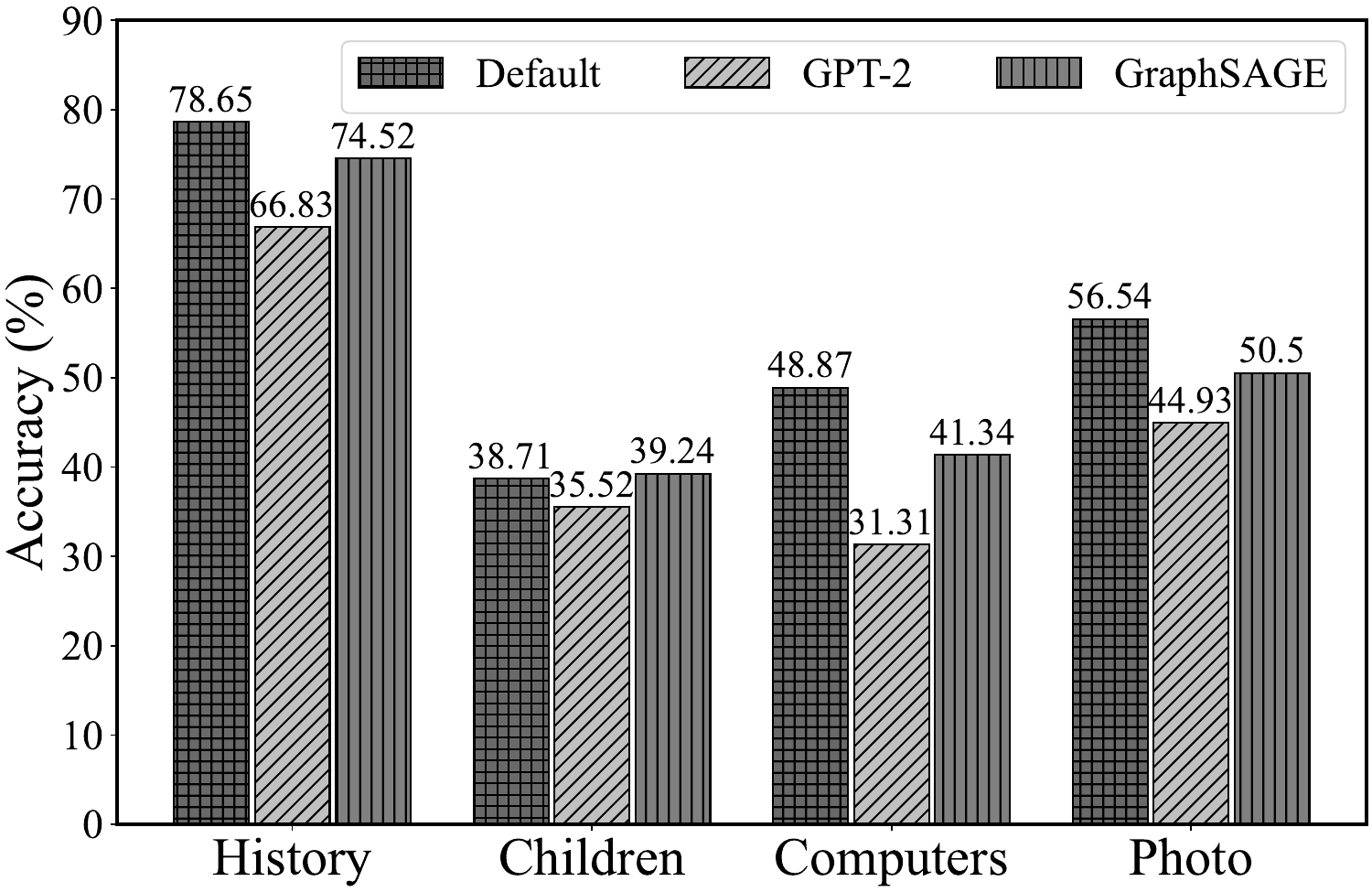}
		\end{minipage}%
	}
	\subfigure[Precision scores of different settings]{
		\begin{minipage}[t]{0.45\linewidth}
			\centering
			\includegraphics[width=1\textwidth]{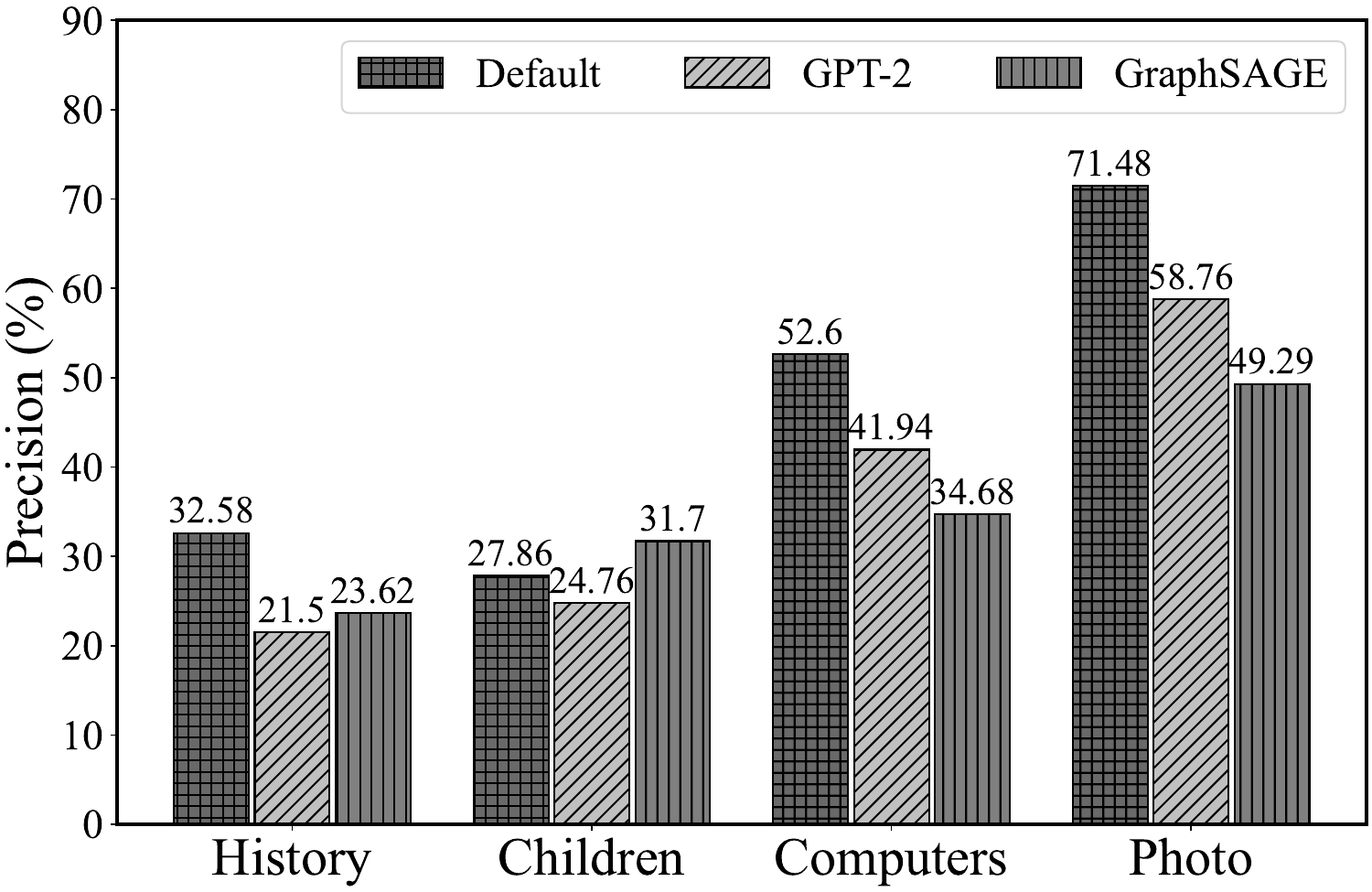}
		\end{minipage}%
	}
	\centering
	\caption{The performance of \name equipped with different text encoder and graph encoder.}
	\label{fig:ablation}
\end{figure*}

\subsubsection{Ablation Study of Text Encoder and Graph Encoder}
\label{sec:adaptor}
There are two critical components in \name: text encoder and graph encoder. In this ablation study, we examined different models for the two components. The text encoder adopts BERT \cite{bert}, and the graph coder is GCN \cite{gcn} in the default settings. Two supplementary experiments are conducted with BERT \cite{bert} being replaced by GPT-2 \cite{gpt-2} and GCN \cite{gcn} being replaced by GraphSAGE \cite{graphsage}, respectively. The experiment results are illustrated in Figure \ref{fig:ablation} below and Figure \ref{fig:app-ablation} in \textbf{Appendix \ref{app:exp}}.

Both BERT and GPT-2 are representative language models in NLP areas. However, there are significant differences between the two models. BERT is a bidirectional model that utilizes tasks like Masked Language Modeling to train word representations, focusing on context-based text understanding. But GPT-2 is a single-direction model trained by self-regression paradigms to predict the next word based on the previous content, which is designed for generative tasks. We can observe that \name equipped with GPT-2 is significantly outperformed by \name equipped with BERT. It indicates that the generative language model is unsuitable for acquiring text feature embeddings. This phenomenon is reasonable as the generative language models are designed for content generation, lacking powerful embedding abilities to obtain informative text representations.

The graph encoder module in \name incorporates the embedded text features and graph structural information to obtain the final representation embedding of the node in the TAG. 
In practice, the graph encoder selected for \name should be simple and efficient for processing large-scale graphs like GCN and GraphSAGE. Though \name equipped with GraphSAGE is functional, it is outperformed by \name equipped with GCN. GraphSAGE is designed for very large graphs and randomly drops some nodes and edges to facilitate the training, which causes information loss.

In short, to ensure the normal functionality and satisfying performance of \name, the text encoder should not adopt generative language models, and the graph encoder should be simple and efficient enough to incorporate the language model to train on large TAGs.

\begin{table}[h]
\centering
\caption{The performance of the proposed \name method with different adaptor settings.}
\label{tab:adaptor}
\renewcommand\arraystretch{1.5}
\resizebox{\textwidth}{!}{
\begin{tabular}{c|cccc|cccc}
\hline\hline
Dataset                        & \multicolumn{4}{c|}{Books-Children}                                       & \multicolumn{4}{c}{Books-History}                                         \\ \hline
\diagbox{Settings}{Metrics} & Accuracy (\%)        & Precision (\%)         & Recall (\%)            & F1 (\%)                & Accuracy (\%)          & Precision (\%)         & Recall (\%)            & F1 (\%)                \\ \hline
Default                        & 38.71 (std 0.65) & 27.86 (std 2.62) & 11.89 (std 0.27) & 12.40 (std 0.43) & 78.65 (std 0.69) & 32.58 (std 4.47) & 25.91 (std 0.77) & 25.55 (std 0.56) \\ \hline
256                            & 40.96 (std 0.51) & 31.19 (std 2.83) & 14.47 (std 0.25) & 15.44 (std 2.75) & 78.86 (std 0.33) & 32.95 (std 3.29) & 26.16 (std 0.67) & 25.75 (std 0.49) \\
512                            & 39.55 (std 0.67) & 28.88 (std 1.85) & 12.73 (std 0.23) & 13.45 (std 0.21) & 79.17 (std 0.43) & 36.33 (std 4.91) & 26.40 (std 0.35) & 25.95 (std 0.22) \\
768                            & 35.28 (std 0.96) & 13.20 (std 2.38) & 8.26 (std 0.33)  & 7.37 (std 0.44)  & 78.48 (std 0.66) & 29.50 (std 4.16) & 25.62 (std 0.93) & 24.98 (std 0.78) \\ \hline\hline
Dataset                        & \multicolumn{4}{c|}{Ele-Computers}                                        & \multicolumn{4}{c}{Ele-Photo}                                             \\ \hline
\diagbox{Settings}{Metrics} & Accuracy (\%)          & Precision (\%)         & Recall (\%)            & F1 (\%)                & Accuracy (\%)         & Precision (\%)         & Recall (\%)            & F1 (\%)                \\ \hline
Default                        & 48.87 (std 0.56) & 52.60 (std 1.62) & 29.48 (std 0.38) & 31.50 (std 0.41) & 56.54 (std 0.40) & 71.48 (std 1.64) & 29.14 (std 0.92) & 35.10 (std 1.31) \\ \hline
256                            & 50.63 (std 0.66) & 53.15 (std 1.34) & 31.22 (std 0.58) & 33.61 (std 0.79) & 57.06 (std 0.51) & 70.94 (std 1.13) & 29.00 (std 0.89) & 34.94 (std 1.14) \\
512                            & 53.44 (std 1.17) & 53.06 (std 1.45) & 34.26 (std 0.95) & 37.01 (std 1.16) & 49.23 (std 0.56) & 49.31 (std 6.85) & 16.58 (std 0.64) & 18.67 (std 0.96) \\
768                            & 48.62 (std 0.30) & 53.59 (std 1.37) & 28.82 (std 0.29) & 30.61 (std 0.40) & 53.86 (std 0.33) & 64.91 (std 5.10) & 24.07 (std 0.70) & 28.96 (std 0.98) \\ \hline\hline
\end{tabular}}
\end{table}

\subsubsection{Adaptor Module Experiment}
As mentioned in Section \ref{sec:text-encoder}, adopting an adaptor module is a common practice for employing pre-trained language models for various downstream applications while avoiding fine-tuning. However, the adaptor module is usually combined with the downstream models to be trained together by supervised signals. But, in our settings, the training phase is motivated by graph contrastive learning, a self-supervised learning paradigm, instead of the supervised one. This section investigates if the adaptor module can apply to \name.

Without losing generality, we employ a single linear layer to decorate the outputs of the text encoder. The adaptor-processed outputs' size is a hyperparameter selected from $\{256, 512, 768\}$. Moreover, the default setting in this experiment denotes the vanilla \name equipped with \textit{shorten} augmentation. The experiment results are shown in Table \ref{tab:adaptor}. 

According to the results, the adaptor module is effective in improving the performance of \name in most scenarios. Specifically, the improvement occurs when the output size of the adaptor is relatively small (\ie smaller than the output size of the text encoder listed in \textbf{Appendix \ref{app:implementation}}). It can be speculated that the role of the adaptor is to condense the text feature embeddings produced by the text encoder to facilitate the following GCL training process.

\section{Related Work}
This section briefly introduces the research works that are highly related to the scope of this paper. The following content is two-fold, which are about LLMs for graph learning and GCL, respectively.

\subsection{Large Language Models for Graph Learning}
LLMs have garnered significant attention for their prowess in natural language processing tasks, but their application in graph learning is a burgeoning field of research \cite{llm-survey, llm-graph-survey}. The intersection of LLMs and graphs presents a promising avenue for enhancing various scientific disciplines such as cheminformatics \cite{cheminformatics}, material informatics \cite{material}, bioinformatics \cite{bio-info}, computer vision \cite{cv}, and quantum computing \cite{quantum}. By incorporating text information with graph data (\ie TAG), researchers can accelerate scientific discovery and analysis, particularly in domains where graphs are paired with critical text properties. A comprehensive survey on LLMs on graphs \cite{llm-graph-survey} categorizes the application scenarios into pure, text-rich, and text-paired graphs, highlighting the diverse contexts in which LLMs can be leveraged. Techniques such as treating LLMs as task predictors \cite{graphformer}, feature encoders for GNNs \cite{tape}, and aligning LLMs with GNNs \cite{glem} offer avenues for exploring the mutual enhancement between LLMs and graphs. However, challenges such as graph linearization, model optimization inefficiencies, and the need for generalizability and robustness of LLMs on graphs underscore the importance of further research in this evolving field \cite{llm-graph-survey}.

\subsection{Graph Contrastive Learning}
The focus of GCL research is on securing high-quality contrasting pairs, which are essential for the effectiveness of GCL. Notable works in the literature have concentrated on creating contrasting pairs through conventional graph augmentation strategies, with satisfying results achieved \cite{dgi, graphcl}. Nonetheless, these approaches have limitations. For example, the randomness inherent in graph augmentation can lead to suboptimal performance in graph-based models \cite{mvgrl}. In response to this challenge, some studies have suggested the creation of various graph views to form contrasting pairs \cite{mvgrl, dsgc} or the adaptive generation of contrasting examples \cite{gca, grace, cgc}. Despite the sophistication of these advanced GCL techniques, they encounter a similar problem to that of the conventional methods: the lack of explicit constraints over the augmentation process. This lack of explicit constraints can result in uncontrollable and incomprehensible outcomes. In contrast, \name leverages an LLM to guide the augmentation of textual attributes by carefully crafted prompts. This approach ensures that both the prompted inputs and the generated outputs are in natural language, offering explicit and comprehensible constraints and results for the augmentation. Furthermore, while existing methods predominantly augment graph structures or feature embeddings, GCL for TAGs is yet to be explored \cite{llm-survey, llm-graph-survey, dataset}. The proposed \name method seeks to extend the reach of GCL techniques to include TAGs, thereby expanding the potential use cases for GCL.

\section{Conclusion}
This paper proposes a novel GCL framework, namely \name, which successfully incorporates LLMs to conduct augmentations to construct contrasting samples. The purpose of the proposed augmentation strategy is to leverage the advantages of LLMs to tackle the limitations of information loss, incapable language models, and implicit constraints of current GCL methods for TAGs, including alleviating information loss during the augmentation, enhancing insufficient NLP abilities of conventional language models, and imposing explicit constraints on the augmentation process. Comprehensive experiments verify the effectiveness and superiority of the proposed \name method. This research is expected to be a pioneering work that encourages the exploration of LLMs for GCL. The future directions are two-fold, including investigating more comprehensive augmentation prompting strategies for different scenarios and how to improve the computation efficiency of employing LLMs in real-world applications.

\begin{ack}
\end{ack}

\bibliographystyle{plain}
\bibliography{ref}

\appendix

\section{Experimental Settings}
The experimental settings, including dataset and baseline descriptions, method implementation details, and evaluation protocols, are listed here to ease the reproducibility of the experiments. More details can be found in the released source codes\footnote{https://anonymous.4open.science/r/LATEX-GCL-0712}.

\subsection{Datasets \& Baselines}
\label{app:data-baselines}
Considering the research scope of this paper, experiments on the graph datasets with promising text attributes are required. Multiple high-quality text-attributed graphs are collected by \cite{dataset} from which four datasets, including Books-Children, Books-History, Ele-Computers, and Ele-Photo, are selected as the experiment datasets. These datasets are extracted from the Amazon dataset \cite{amazon-1, amazon-2}, which have raw text descriptions for each node and are large-scale compared to previous text-attributed graph datasets \cite{dataset}. The statistics and the content of the raw text of each dataset are listed in Table \ref{tab:dataset}.

Besides the datasets, five impactful GCL methods are selected as baselines for the comparison study. These baselines can be roughly broken down into three categories: I) GraphCL \cite{graphcl} is the most classical GCL method that involves several conventional random-based augmentations, II) GCA \cite{gca} and GRACE \cite{grace} are both the adaptive augmentation-based GCL methods, where GCA conducts automatic selection from the conventional augmentation techniques and GRACE performs trainable augmentations based on the input graph data, and III) both BGRL \cite{bgrl} and GBT \cite{gbt} method follow a novel GCL paradigm that utilizes different training objectives instead of InfoNCE \cite{infonce} based on DGI \cite{dgi} to eliminate the requirement of negative contrasting samples to achieve storage efficient.

\subsection{Method Implementation Details}
\label{app:implementation}
The LLM used for dataset augmentations in our settings is \textit{GPT-3.5-turbo}, and the specific version is default and decided by OpenAI update schedule\footnote{https://platform.openai.com/docs/models/gpt-3-5-turbo}. The prompts for guiding the LLM to generate augmented text are listed in Table \ref{tab:aug} in the methodology section.

Moreover, we adopt a pre-trained BERT \cite{bert} model, whose version is \textit{bert-base-uncased}, to embed the original and augmented text attributes. The pre-trained model and other related components are used according to the guidance of \textit{Pytorch-Transformers}\footnote{https:\/\/pytorch.org\/hub/huggingface\_pytorch-transformers\/}. The pre-trained model and other related components can be publicly accessed on \textit{Hugging Face} via this link\footnote{https://huggingface.co/google-bert/bert-base-uncased}.

Some important hyperparameter settings are listed here. The embedding size of the text encoder is set to 768, and the output size of the graph encoder is set to 256. The learning rate for the whole framework training is $2e^{-5}$. The training batch size and the epoch number are set to 512 and 10, respectively. For more implementation details, please refer to the released source codes. All the experiments are performed on NVIDIA RTX A5000 24GB.

\subsection{Evaluation Protocol}
\label{app:evaluation}
The proposed method is evaluated based on the node classification task, which is subject to the linear evaluation protocol. The linear evaluation is to train and test a support vector machine (SVM) on node feature embeddings trained by the method to be evaluated to verify the quality of the outputs of the proposed \name method, where the SVM is implemented by a third-party toolkit named \textit{scikit-learn}\footnote{https://scikit-learn.org/}. Specifically, to ensure the reliability of the experiment results, we repeat the experiment five times. For each time, 20\% of the nodes are selected as the training set, and 10\% of the rest of the nodes are the test set. Sufficient metrics, including Accuracy, Precision, Recall, and F1 scores with standard deviations, are used to demonstrate the results of the linear evaluation.

\section{Supplementary Experiment Results}
\label{app:exp}
\begin{figure*}[h]
	\centering
        \subfigure[Recall scores of different settings]{
		\begin{minipage}[t]{0.45\linewidth}
			\centering
			\includegraphics[width=1\textwidth]{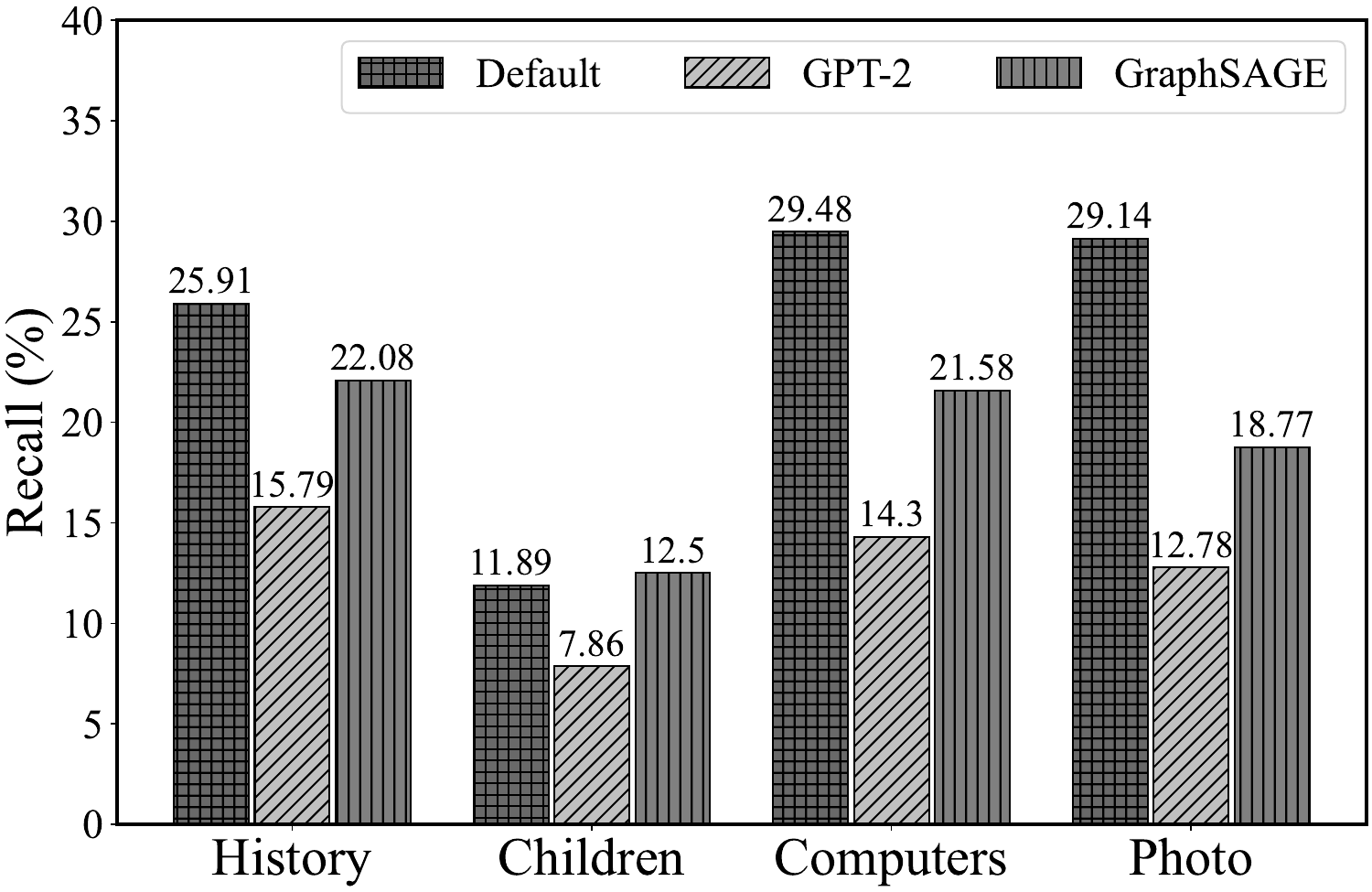}
		\end{minipage}%
	}
        \subfigure[F1 scores of different settings]{
		\begin{minipage}[t]{0.45\linewidth}
			\centering
			\includegraphics[width=1\textwidth]{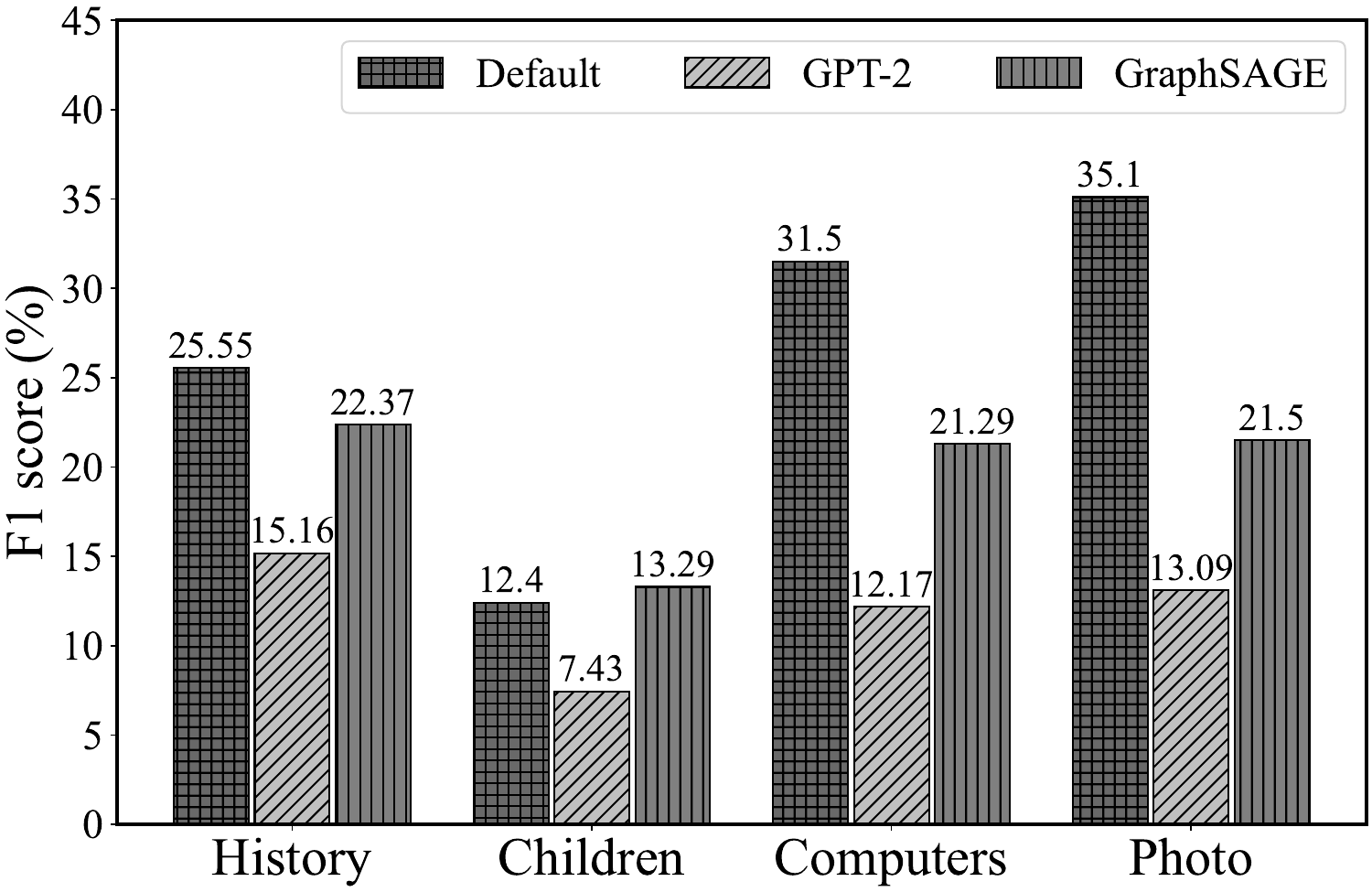}
		\end{minipage}%
	}
	\centering
	\caption{The performance of \name equipped with different text encoder and graph encoder.}
	\label{fig:app-ablation}
\end{figure*}

The performance of LATEX-GCL equipped with different text encoder and graph encoder measured by metrics Recall and F1 are shown in Figure \ref{fig:app-ablation}

\end{document}